\documentclass[
reprint,
showpacs,
nofootinbib,
amsmath,amssymb,
aps,
pra,
floatfix,longbibliography
]{revtex4-2}
\usepackage{graphicx}
\usepackage{dcolumn}
\usepackage[hypertexnames, breaklinks=true, bookmarksnumbered=true, bookmarksopen=true, colorlinks=true, linktocpage=true, citecolor=blue, urlcolor=blue,linkcolor=red]{hyperref}
\usepackage{amsfonts,amsthm}
\usepackage{bbm} 
\usepackage{bm} 
\usepackage{mathtools} 
\usepackage[stretch=10]{microtype} 
\usepackage{empheq}
\usepackage{color}
\usepackage{tcolorbox}
\usepackage[T1]{fontenc}
\usepackage[latin9]{inputenc}
\usepackage{txfonts}
\usepackage{tikz}
\usetikzlibrary{matrix, arrows.meta, decorations.pathreplacing}
\usepackage{natbib}
\allowdisplaybreaks

\newcommand{\eeqref}{Eq.~\eqref}
\newcommand{\ccite}{Ref.~\cite}

\newcommand{\CC}{\mathbbm{C}}

\newcommand{\MM}{\mathbbm{M}}

\newcommand{\U}{\mathcal{U}}

\DeclareMathOperator{\tr}{Tr}

\newcommand{\idm}{\mathbbm{I}}

\newcommand{\bra}[1]{\langle #1 |}
\newcommand{\ket}[1]{| #1 \rangle}
\newcommand{\braket}[2]{\langle #1 | #2\rangle}

\newtheorem{lemma}{\emph{Lemma}}

\newtheorem{defn}{Definition}

\newtheorem{rmk}[defn]{Remark}

\newtheorem{thm}[defn]{Theorem}

\theoremstyle{remark}

\newtheorem*{rmk*}{Remark}

\begin{document}
	
		
	\title{ Unital qubit channels are $3$-noisy}
	\author{Indu Bala}
	\email{indu.qic@gmail.com}
	\author{Preeti Parashar}
		\email{parashar@isical.ac.in}
	\author{Swapan Rana}
		\email{swapanqic@gmail.com}
	\affiliation{Physics and Applied Mathematics Unit, Indian Statistical Institute, 203 B. T. Road, Kolkata 700108, India}
	\date{\today}
	
	\begin{abstract} 
		The algebraic and environmental characteristics of mixed unitary quantum operations are investigated. It is established that the composition of two uniformly n-mixed unitary channels is itself n-noisy, and this n-noisy property is shown to be an invariant under unitary equivalence. Furthermore, by utilizing majorization and the divisibility of Kraus rank-four unital qubit channels, we demonstrate that any unital qubit channel can be physically simulated using a three-dimensional maximally mixed environment. Together, these results advance the classification and physical realization of noisy quantum channels.
		\end{abstract}
	\maketitle
	\section{Introduction}
	Quantum channels \cite{CHOI1975285} are a fundamental component in quantum physics for modeling how a quantum system evolves over time. The Stinespring dilation theorem dictates that any quantum channel $\Lambda$ can be represented as a joint unitary evolution between the primary system and an external environment \cite{Stinespring}. Since a $d$-dimensional channel possesses at most $d^2$ Kraus operators, it can always be implemented using a $d^2$-dimensional pure environment state. 
	Relying on pure ancillas posses significant practical and physical challenges. Experimentally, preparing and maintaining a high dimensional pure state requires costly active cooling, active reset operation and precise state initialization. A natural question then arises: can we physically simulate a quantum operation using an environment dimension strictly less than $d^2$? 
	
	Terhal \textit{et al.} \cite{terhal1999simulating} demonstrated that a mixed qutrit environment suffices to simulate the generalized depolarizing channel, and subsequently conjectured that any qubit channel could be simulated using a qutrit environment. The simulation of qubit channels has since been extensively discussed in various different contexts \cite{narang2007simulating, abu2026detecting, Karol, Benjamin, Lloyd}. In realistic open-system settings, an external bath is rarely in a pristine zero-temperature state; rather, it typically represents a high-temperature, randomized environment about which we possess zero prior information \cite[Chapter 3]{Peter}. Coupling a quantum system to such a maximally mixed bath ensures that the environment maintains no preferred energetic bias---preventing it from artificially pumping energy into or draining energy out of the system \cite{mendl}. To formally model such interactions at infinite-temperature, unpolarized baths, Horodecki et al. \cite{Horodecki} originally introduced the concept of $n$-noisy operations in the context of quantum thermodynamics.
	A quantum channel $\Lambda : \mathbb{M}_d(\mathbb{C}) \to \mathbb{M}_d(\mathbb{C})$ is defined as $n$-noisy if it can be realized via an $n$-dimensional maximally mixed environment:
	\begin{equation}\label{eq:n-noisy}
		\Lambda(\rho) = \operatorname{Tr}_{E}\left( U \left( \rho \otimes \frac{I_n}{n} \right) U^\dagger \right),
	\end{equation}
	where $U \in \mathcal{U}(\mathbb{C}^d \otimes \mathbb{C}^n)$. From Eq.~\eqref{eq:n-noisy}, it is immediate that any $n$-noisy channel must be unital. More broadly, such operations fall under the framework of factorizable quantum channels \cite{Haagerup2011, Haagerup_2015, Levick2018, Haagerup2021,musat2025}. While every mixed-unitary channel is factorizable, the converse is not true in general. 
	
	Interestingly, there exist qutrit mixed-unitary channels that fail to be $n$-noisy for any $n \in \mathbb{N}$ \cite{muller2019}. Müller-Hermes and Perry \cite{muller2019} established that every unital qubit channel is $4$-noisy. For single-qubit systems ($d = 2$), the central problem reduces to determining the minimal environment dimension $n$ required to simulate an arbitrary unital channel. Environmental dimension cannot be reduced upto 2 as Kraus rank 3 qubit channel  cannot be $2$-noisy, as a Schmidt rank of 3 is impossible for a joint unitary operator $U \in \mathcal{U}(\mathbb{C}^2 \otimes \mathbb{C}^2)$ \cite{Ion}. Consequently, the authors of \ccite{muller2019} posed the open question of whether any unital qubit channel can be universally simulated using a $3$-noisy environment.
	 While this problem has previously remained open for the Kraus rank-4 case, in this work we demonstrate that an environment dimension of 3 is optimal by exploiting the framework of channel divisibility.
The article is structured as follows: Section~\ref{sec:divisible} examines quantum channel divisibility, and Section~\ref{sec:invar} addresses the invariance of $n-$ noisy channels. Concluding remarks and future research directions are presented in Section~\ref{sec:conc}.

\section{Divisibility of quantum channels}\label{sec:divisible}
Since channel divisibility is closely linked to channel implementation \cite{PRABidivisible} and can be utilized to realize quantum operations with smaller registers, we discuss this framework in further detail. A quantum channel $\Lambda$ is said to be divisible if it can be written as the concatenation of two non-unitary channels, i.e.,
\begin{equation}
	\Lambda=\Lambda_1\circ \Lambda_2.
	\label{eq:divisibility}
\end{equation}
The concept of channel divisibility was originally introduced by Wolf \textit{et al.} \cite{wolfdividing}, who provided a complete characterization of the sets of indivisible and infinitesimally divisible qubit channels. Notably, every $d$-dimensional quantum channel of maximum Kraus rank $d^2$ is divisible \cite[Theorem~11]{wolfdividing}.  If $T$ is the matrix representation of a divisible quantum channel $\Lambda=\Lambda_1\circ\Lambda_2$, its matrix can be decomposed as
\begin{equation}	\label{eq:matrix_decomp}
	T=T_1T_2,
\end{equation}
where $T_1$ is an invertible matrix representing a channel sufficiently close to the identity operation. In this way, we apply this framework to decompose Pauli channels of Kraus rank 4.

\subsection{Divisibility of unital qubit channels}
Because Kraus rank-1 channels correspond to unitary operations, they are by definition trivially indivisible. Therefore, non-trivial divisibility only applies to channels with a Kraus rank greater than or equal to 2. The complete decomposition of Kraus rank-2 qubit channels was established in Ref.~\cite{wolfdividing}, while Kraus rank-3 unital qubit channels are known to be indivisible. In this section, we first decompose a Kraus rank-4 unital qubit channel into a concatenation of a Kraus rank-3 and a Kraus rank-2 Pauli channel. Subsequently, we express it as a combination of two channels, each comprising an equal mixture of three unitaries.

	\begin{thm}\label{thm-rank4div}
		Every unital qubit channel of Kraus rank 4 can be written as a composition of two unital channels, one of Kraus rank $3$ and the other of Kraus rank $2$.
	\end{thm}
	\begin{proof}
		Up to a unitary equivalence, any Kraus rank 4 unital qubit channel can be written as \cite[Theorem~2.2]{CHOI1975285}
		\begin{equation}\label{Eq:Gen.KR4.UQC}
		\Lambda(X)=\sum_{i=0}^3 p_i\,\sigma_i X \sigma_i^\dagger,\quad 0<p_{i+1}\leq p_i.
		\end{equation}

	For the channel in \eeqref{Eq:Gen.KR4.UQC}, the channel matrix $T$ \cite{BETHRUSKAI2002159} is given by  
	\begin{equation}\label{Eq:Channel.T}T=\begin{pmatrix}
		1 & 0 & 0 & 0\\
		0 & \lambda_1 & 0 & 0\\
		0 & 0 &  \lambda_2 & 0\\
		0 & 0 &0&   \lambda_3 
	\end{pmatrix},
	\end{equation}
	where the $\lambda_i$'s are given by $\lambda_1=p_0-p_1+p_2-p_3,\,\lambda_2=p_0-p_1-p_2+p_3,\,\lambda_3=2(p_0+p_1)-1.$ Inverting $\lambda$ into $p$, we get
	\begin{equation}\label{Eq:Pi.Lambdai}
	\begin{split}
	p_0&=\frac{1}{4}\left(1+\lambda_1+\lambda_2+\lambda_3\right),\,
	p_1=\frac{1}{4}\left(1-\lambda_1-\lambda_2+\lambda_3\right),\\
	p_2&=\frac{1}{4}\left(1+\lambda_1-\lambda_2-\lambda_3\right),\,
	p_3=\frac{1}{4}\left(1-\lambda_1+\lambda_2-\lambda_3\right).\\
	\end{split}
	\end{equation}
	As every Kraus rank $4$ qubit channel is divisible, and divisibility remains invariant under unitary equivalence \cite{wolfdividing}, we can decompose $T$ of \eeqref{Eq:Channel.T} in the following way
	\begin{equation}\label{Eq:T=T1.T2}
		T=T_1T_2:=\begin{pmatrix}
		1 & 0 & 0 & 0\\
		0 & \frac{\lambda_1}{t}& 0 & 0\\
		0 & 0 & \lambda_2 & 0\\
		0 & 0 & 0 & \frac{\lambda_3}{t}\\
	\end{pmatrix}\begin{pmatrix}
		1 & 0 & 0 & 0\\
		0 & t & 0 & 0\\
		0 & 0 &  1 & 0\\
		0 & 0 &0&  t 
	\end{pmatrix}
	\end{equation}
	for some $t\neq 0$, whose value(s) would be determined from the channel conditions (complete positivity). From \eeqref{Eq:Channel.T} and \eeqref{Eq:Pi.Lambdai}, $T_2$ corresponds a channel iff $|t|\leq 1$.
	
	Since a unital channel of Kraus rank 3 is indivisible \cite[Theorem~23]{wolfdividing}, $\Lambda_1$  is indivisible if one of the eigenvalue of $\Lambda_1$ is zero. So we can take \[1-\frac{\lambda_3+\lambda_1}{t}+\lambda_2=0 \]
	\[\implies t=\frac{\lambda_3+\lambda_1}{1+\lambda_2}=\frac{p_0-p_3}{p_0+p_3}.\]
	Now we will discuss the following two cases.
	
	\textbf{Case I: $t\neq 0$}, i.e., $p_0>p_3$.
	
	One verifies that $\Lambda=\Lambda_1\circ\Lambda_2$, 
\begin{equation}\label{eq:Phi:I}
	\begin{split}
		\Lambda_2(\rho)&=\frac{1+t}{2}\rho+\frac{1-t}{2}Z\rho Z=\frac{1}{p_0+p_3}\left(p_0\rho+p_3Z\rho Z\right)\\
	\Lambda_1(\rho)&=	a\rho+bX\rho X+cY\rho Y,
	\end{split}
	\end{equation}
	where \
	\begin{equation}\label{eq:abc}
	\begin{split}
		a&= \frac{1}{4}\left(1+\lambda_2+\frac{\lambda_3+\lambda_1}{t}\right)=p_0+p_3\\
		b&=	\frac{1}{4}\left(1-\lambda_2+\frac{\lambda_3-\lambda_1}{t}\right)=	\frac{p_0p_1-p_3p_2}{p_0-p_3}\\
		c&=	\frac{1}{4}\left(1-\lambda_2-\frac{(\lambda_3-\lambda_1)}{t}\right)=	\frac{p_0p_2-p_3p_1}{p_0-p_3}.
	\end{split}
	\end{equation}
	
	\textbf{Case II: $t=0$,} i.e., $p_0=p_3$ $\Rightarrow p_i=\frac{1}{4}\,\forall i=0,\hdots, 3$. 
	
	In this case, our channel will be the completely depolarizing channel. One verifies that
 \[T=\begin{pmatrix}
	1 & 0 & 0 & 0\\
	0 & 0 & 0 & 0\\
	0 & 0 & 0 & 0\\
	0 & 0 &0&  0 
\end{pmatrix}=T_1T_2=\begin{pmatrix}
	1 & 0 & 0 & 0\\
	0 & \frac{1}{2}& 0 & 0\\
	0 & 0 & 0 & 0\\
	0 & 0 & 0 & \frac{1}{2}\\
\end{pmatrix}\begin{pmatrix}
	1 & 0 & 0 & 0\\
	0 & 0 & 0 & 0\\
	0 & 0 &  1 & 0\\
	0 & 0 &0&  0 
\end{pmatrix}\]

\begin{equation}
	\begin{split}
	\Lambda_1(\rho)
	&=
	\frac{	1}{2}\rho+\frac{1}{4}X\rho X+\frac{1}{4}Y\rho Y,
	\label{eq:PhiA}
	\\
	\Lambda_2(\rho)
	&=
	\frac{1}{2}\rho+\frac{1}{2}Z\rho Z.
	\end{split}
\end{equation}
\end{proof}

We will now show that each of the factor channels $\Lambda_1$ and $\Lambda_2$ can be taken to be a uniform mixture of three unitaries. For that we will use the following result.

\begin{lemma}\cite{choi2023unital}\label{lemma-equalmixture}
	A unital qubit channel can be expressed as a convex combination of unitary channels with convex coefficients $p:=(p_1,\dotsc,p_m)$ as long as $p \prec \mu$, $\mu$ being the vector of eigenvalues of the Choi state of the channel.
\end{lemma}
\begin{thm}\label{thm-mixture-unitary}
	Every Kraus rank 4 unital qubit channel can be written as a composition of two channels, each of which is an equal mixture of 3 unitary channels.
\end{thm}
\begin{proof}
As in the proof of Theorem~\ref{thm-rank4div}, W.L.O.G. we can take $\Lambda_{p}$ as the channel in \eeqref{Eq:Gen.KR4.UQC}. From Theorem~\ref{thm-rank4div}, we can write 
\[\Lambda_p=\Lambda_1\circ\Lambda_2.\]
It will suffice to show that in each cases, the coefficients appearing in the factors of $\Lambda_i$ will  majorize  the uniform vector $(1/3,1/3,1/3)$. 

\textbf{Case~I: $p_0>p_3$}, $\Lambda_1,\Lambda_2 $ are given by Eqs.~ \eqref{eq:Phi:I} and \eqref{eq:abc}.

From \eqref{eq:Phi:I}
	we can verify 
\[1=p_0+p_1+p_2+p_3\leq3p_0+p_3\leq3(p_0+p_3),\]
and hence
\begin{equation}\label{eq-a}
	a\geq\frac{1}{3}.
\end{equation}
We also have
\[c-p_2=\frac{p_3(p_2-p_1)}{p_0-p_3}\leq0.\]

$p_2\leq\frac13$ implies $c\leq\frac{1}{3}$, it follows 
\begin{equation}\label{eq-a+b}
a+b\geq\frac{2}{3}.
\end{equation}
From  Eqs.~\eqref{eq-a} and \eqref{eq-a+b} it follows that
\[(a, b, c) \succ \left( \frac{1}{3}, \frac{1}{3}, \frac{1}{3} \right).\]

So, by Lemma~\eqref{lemma-equalmixture},  there exiat unitaries $V_i$  such that 
\[\Lambda_1(\rho)= \frac{1}{3}\sum_iV_i\rho V_i^\dagger.\]
Now from \eeqref{eq:Phi:I}, \[\Lambda_2(\rho)=\frac{1}{p_0+p_3}\left(p_0\rho+p_3Z\rho Z\right),\]
\[\text{where }\quad\frac{p_0}{p_0+p_3}\geq\frac{1}{2}\geq\frac{1}{3}.\]
Therefore, 
\[\left(\frac{p_0}{p_0+p_3}, \frac{p_3}{p_0+p_3}, 0\right) \succ \left( \frac{1}{3}, \frac{1}{3}, \frac{1}{3} \right)\]
Hence there will exist unitaries $\{W_i\}_{i=1}^3$ such that 
\[\Lambda_2(\rho)=\frac{1}{3}\sum_{i=1}^3W_iXW_i^\dagger.\]
Thus, for $p_0>p_3$ there exist channels $\Lambda_1,\Lambda_2$ which are an equal mixture of three unitary channels such that \[\Lambda_{p_0>p_3}=\Lambda_1\circ\Lambda_2.\]

\textbf{Case~II: $p_0=p_3$}. Since \[\left(\frac{1}{2},\frac{1}{4}, \frac{1}{4}\right) \succ \left( \frac{1}{3}, \frac{1}{3}, \frac{1}{3} \right)\,\text{and } \left(\frac{1}{2},\frac{1}{2}, 0\right) \succ \left( \frac{1}{3}, \frac{1}{3}, \frac{1}{3} \right).\]
Thus, in this case also we can write 
$\Lambda=\Lambda_1\circ \Lambda_2$, where $\Lambda_1$ and $\Lambda_2$ are an equal mixture of three unitary channels. For completeness, an explicit form of the unitaries $V_i,W_i$ is given in Appendix.
\end{proof}
		
\section{Invariance properties and 3-noisy simulation\label{sec:invar}}

		In this section, we establish the core algebraic properties of $n$-noisy quantum operations. Specifically, we prove that the property of being $n$-noisy is invariant under unitary equivalence, and that the composition of two uniformly $n$-mixed unitary channels remains $n$-noisy. Leveraging these structural theorems alongside channel divisibility, we ultimately conclude that any unital qubit channel can be physically simulated using a 3-dimensional maximally mixed environment.
		
\begin{thm}\label{thm-UE}
		If channel  $\Lambda$ and $\Psi$ are  unitarily equivalent then $\Lambda$ is $n$-noisy if and only if $\Psi$ is n-noisy
	\end{thm}
	\begin{proof}
		Since channel $\Lambda$ is unitarily equivalent to $\Psi$, there will exist unitary R and S such that
		\[\Lambda[\rho]=R\,\Psi\left[S\,\rho\,S^\dagger\right]R^\dagger.\]
		If we assume $\Psi:\MM_d(\CC)\to \MM_d(\CC)$	 is n-noisy so there will exist unitary $U\in \U(dn)$ such that
		\begin{equation}\label{eq-implement U}
		\Psi(\rho)=\tr_{2}\left[U\left(\rho\otimes\frac{I_n}{n}\right)U^\dagger\right].
		\end{equation}
		Defining $\widetilde U=(R\otimes I_n)U(S\otimes I_n)$, we note that $\widetilde U\in \U(dn)$. Moreover,
		\begin{align*}
			&\tr_{2}
			\left[
			\widetilde U
			\left(
			\rho\otimes\frac{I_n}{n}
			\right)
			\widetilde U^\dagger
			\right]\\
			&\qquad=
			R\,
			\Psi
			\left(
			S\rho S^\dagger
			\right)
			R^\dagger\\
			&\qquad=
			\Lambda(\rho).
		\end{align*}
		Thus, the original channel is also n-noisy.
	\end{proof}
	\begin{rmk}
		Every unital qubit channel is unitarily equivalent to a Pauli channel \cite[Theorem 2.2]{choi2023unital}, so by Theorem~\ref{thm-UE}, we only need to check the noisiness of Pauli channels.
	\end{rmk}
	If a unital channel can be expressed uniform mixtures of $n$ number of unitray channels, it need not be $n$-noisy. Counterexample: Kraus rank $3$ Pauli Channel \cite[Theorem 3.2]{muller2019}. But if  we have two channels each of which is an equal mixture of $n$ unitary channels, then their composition will always be $n$- noisy, as given by the following result.  
	\begin{thm}\label{thm-nnoisy-comp}
		The composition $\Lambda_1 \circ \Lambda_2$ of two channels that are uniform mixtures of $n$- unitaries is itself $n$-noisy.
	\end{thm}	
	\begin{proof}
		Let us assume $\Lambda_1,\Lambda_2:\MM_d(\CC)\to\MM_d(\CC)$  are in equal mixture of $n$ unitary channels, i.e.
		\begin{align}\label{eq-VW}
			\begin{split}
		\Lambda_1(X)&=\frac{1}{n}\sum_{i=1}^nV_iXV_i^\dagger,\\
        \Lambda_2(X)&=\frac{1}{n}\sum_{i=1}^nW_iXW_i^\dagger
        \end{split}
	\end{align}
		where $V_i$ and $W_i$ are the unitaries.
		Next we will connect the unitaries $\{V_i,W_i\}$ in  Eq~\eqref{eq-VW} using the Fourier unitary F and will get another unitary U which implement $\Lambda_1\circ\Lambda_2$ via Eq~\eqref{eq-implement U}. 
		
\begin{equation}\label{eq~U}
			U=\sum_{j,k=1}^nF_{j,k}V_jW_k \otimes \ket{j}\bra{k},\, F_{j,k}= \frac{1}{\sqrt{n}} e^{-\frac{2\pi i}{n} (j-1)(k-1)}.
		\end{equation}
	\begin{align*}
		\begin{split}
		U^\dagger U&=\sum\limits_{j,k,l,m=1}^n \overline{F_{j,k}}F_{l,m}W_k^\dagger V_j^\dagger V_lW_m\otimes \ket{k}\braket{j}{l}\bra{m},\\
		&=\sum\limits_{j,k,m=1}^n \overline{F_{j,k}}F_{j,m}W_k^\dagger W_m\otimes \ket{k}\bra{m},\\
		&=\sum\limits_{k,m=1}^n\delta_{k,m}W_k^\dagger W_m\otimes \ket{k}\bra{m} =\idm.
		\end{split}
		\end{align*}
	So, the operator $U$ in Eq~\eqref{eq~U} is indeed a unitary. Next, it follows that
 \begin{align*}
 	&\tr_{2}\left(U\left(\rho\otimes\frac{I_n}{n}\right)U^\dagger\right)=\frac{1}{n}\sum\limits_{j,k=1}^nF_{j,k}\overline{F_{j,k}}V_jW_k\rho W_k^\dagger V_j^\dagger,\\
		&=\frac{1}{n^2}\sum\limits_{j,k=1}^nV_jW_k\rho W_k^\dagger V_j^\dagger=\Lambda_1\circ\Lambda_2.
		\end{align*}
	Hence $\Lambda_1\circ\Lambda_2$ is $n$- noisy.
	\end{proof}

	\begin{thm}\label{thm-comp}
		Unital qubit channels are $3$-noisy.
	\end{thm}
	\begin{proof}
	If a unital qubit channel has Kraus rank $\leq 3$, it will always be $3$-noisy by Ref.~\cite[Corollary 2.4]{muller2019}. So only remaining case to prove  is Kraus rank $4$. From Theorem~\eqref{thm-mixture-unitary} for any unital qubit channel $\Lambda$ of Kraus rank $4$  there exist  $\Lambda_1$ and $\Lambda_2$ channels which are in  equal mixture of three unitary channel such that \[\Lambda=\Lambda_1\circ\Lambda_2.\]  Then,  by Theorem~\ref{thm-nnoisy-comp},  $\Lambda=\Lambda_1\circ\Lambda_2$ is 3-noisy.  \\
	\end{proof}
	
\section{Conclusion}\label{sec:conc}
In this work, we have established that the property of being an $n$-noisy quantum channel is strictly preserved under both channel composition and unitary equivalence. Furthermore, by exploiting the framework of channel divisibility and majorization, we have demonstrated that a three-dimensional maximally mixed environment is universally sufficient to physically simulate any unital qubit channel. 

For future research, it would be of significant interest to explore the precise operational connections between channel divisibility and $n$-noisy implementations in higher-dimensional quantum systems. Investigating whether similar environmental bounds can be derived for higher-dimensional unital channels remains an open and promising avenue for advancing the classification of noisy quantum dynamics.
\bibliographystyle{apsrev4-2}
\bibliography{3n}
\section*{Apppendix: Explicit form of $V_i, W_i$ in Theorem~\ref{thm-mixture-unitary}}
One explicit form for the unitaries $V_i$ and $W_i$ in Theorem~\ref{thm-mixture-unitary} are given by the following expression.
\begin{equation}\label{eq:V_i}
	\begin{split}
	V_1&=\alpha I+i\epsilon Y\\
V_2&=-\beta I+i \gamma X+i\delta Y\\
	V_3&=-\beta I-i\gamma X+i\delta Y.
\end{split}
\end{equation}
\begin{equation}\label{eq:W_i}
	\begin{split}
		W_1&=I,\\
		W_2&=rI+isZ	\\
		W_3&=rI-isZ.
	\end{split}
\end{equation}
\begin{equation}\label{eq:parameter}
	\begin{split}
		\alpha& =\sqrt{3at},\quad \beta=\sqrt{\frac{3au}{2}},\quad\gamma =\sqrt{\frac{3b}{2}},\quad
		\delta=\sqrt{\frac{3ct}{2}},\\\epsilon&=\sqrt{3cu},
		 r=\sqrt{1-\frac{3d}{2}},\quad s=\sqrt{\frac{3d}{2}},\quad d=\frac{p_3}{p_0+p_3}
	\end{split}
\end{equation}
In the first case, when   $p_0>p_3$, the coefficients $a,b,c$  are given by \eeqref{eq:abc} and 
 $t=\frac{\frac13-c}{a-c},\quad	u=1-t$. And for the second case, i.e. when $p_0=p_3$, we choose
 $a=d=t=u=\frac{1}{2}$ and $b=c=\frac{1}{4}$.  	
To prove that it will recover our original channel back, we will show that the Choi matrix of both the channels are same. In our case,  \[\Lambda(\rho)=\Lambda_1\circ\Lambda_2=\frac{1}{9}\sum_{i,j=1}^3 V_iW_j\,\rho\, W_j^\dagger V_i^\dagger.\]
Evidently, the Choi states are 
\begin{align*}
J_{\Lambda}&=\frac{1}{2}\sum_{i,j}E_{i,j}\otimes \Lambda\left(E_{i,j}\right)\\
&=\frac{1}{2}\begin{pmatrix}
	p_0+p_3 & 0 & 0 &p_0-p_3\\
	0 & p_1+p_2 &p_1-p_2 &0\\
	0 & p_1-p_2& p_1+p-2 & 0\\
	p-0-p_3 & 0&0& p-0+p_3
\end{pmatrix}\\
&=J_p.
\end{align*}

\end{document}